\documentclass[11pt, A4paper]{article}
        \usepackage{amsmath}
        \usepackage{amssymb}
        \usepackage{graphicx}

        \newcommand{\ve}[1]{\mbox{\boldmath$#1$}}
        \newcommand{\muas}[0]{\hbox{\rm $\mu$as}}

\begin{document}
	
	\setcounter{figure}{0}
	\setcounter{table}{0}
	\setcounter{footnote}{0}
	\setcounter{equation}{0}
	
	\noindent {\Large\bf TOTAL LIGHT DEFLECTION IN THE GRAVITATIONAL FIELD OF SOLAR SYSTEM BODIES}
	\vspace*{0.7cm}

	\noindent\hspace*{1cm} S. ZSCHOCKE\\[0.2cm]
	\noindent\hspace*{1cm} TU Dresden, Institute of Planetary Geodesy - Germany - sven.zschocke@tu-dresden.de\\
	
	\vspace*{1cm}
	
	\noindent {\large\bf ABSTRACT.} 
	The total light deflection represents a concept, which allows one to decide which multipoles need to be implemented in the light trajectory for a given astrometric accuracy. 
	The fundamental quantity of total light deflection is the tangent vector of the light trajectory at future infinity. It has been found that this tangent vector is naturally given 
	by Chebyshev polynomials. It is just this remarkable fact, which allows to determine strict upper limits of total light deflection for each individual multipole 
	of solar system bodies. Special care is taken about the gauge terms. It is found that these gauge terms vanish at spatial infinity. The results are applied to the case of 
        light deflection in the gravitational fields of Jupiter and Saturn.
	
	\vspace*{1cm}
	
	\noindent {\large\bf 1. INTRODUCTION}
	\smallskip
	
Angular measurements of stellar objects have made impressive advancements during recent decades. In particular, the astrometry missions {\it Hipparcos} 
and {\it Gaia} of European Space Agency (ESA) have reached the milli-arcsecond (mas) and the micro-arcsecond (\muas) level of accuracy, respectively. 
The next goal in astrometric science is to arrive at the sub-micro-arcsecond (sub-$\mu{\rm as}$) or even the 
nano-arcsecond (nas) scale of accuracy. The objectives of such highly precise measurements are overwhelming, e.g.: detection of earth-like planets, 
stringent tests of relativity, mapping of dark matter from areas beyond the Milky Way, and direct distance measurements of stellar standard candles
up to the closest galaxy clusters; see also (Johnston, 2000). 

In fact, several missions have been proposed to ESA, aiming at such levels in astrometric precision, like {\it Theia} and {\it Gaia-NIR}, which are primarily designed 
to study local dark matter properties, to detect Earth-like exoplanets, and to study the physics of highly compact objects (white dwarfs, neutron stars, black holes). 
A further promising candidate is {\it NEAT} (Near Infrared Astrometric Telescope), originally designed for an precision of about 50 nas. 

The fundamental assignment in relativistic astrometry is the precise interpretation of observational data, which requires 
an accurate modeling of trajectories of light signals through the curved space-time of the solar system. In view of recent achievements 
in astrometric angular observations as well as in view of missions proposed to ESA, a corresponding development in the theory of light propagation 
is indispensable. The investigation of the total light deflection is a further step towards these directions. 
 
	
	\vspace*{0.7cm}
	\noindent {\large\bf 2. THE METRIC TENSOR}
	\smallskip

The curved space-time is described by the pair $\left({\cal M}, g_{\mu\nu}\right)$ where ${\cal M}$ is a four-dimensional differentiable manifold, while $g_{\mu\nu}$ is 
the metric tensor of the manifold, and each point ${\cal P} \in {\cal M}$ represents a space-time event. The metric tensor is governed by the field equations of 
gravity (Einstein, 1915), which relate the metric tensor $g_{\alpha\beta}$ of the physical manifold ${\cal M}$ to the stress-energy tensor of matter $T_{\alpha\beta}$.  
These exact field equations can only be solved in closed form for highly symmetric bodies, like spherically symmetric bodies or bodies of ellipsoidal shape,
but not for realistic bodies of the solar system. Therefore, approximative
approaches of general relativity are essential for further progress in the theory of gravity and in the theory of light propagation.
In the solar system the gravitational fields are weak and, therefore, one may apply the theory of linearized gravity. In that approximation, the covariant components 
of the metric tensor are decomposed into the flat Minkowski metric $\eta_{\alpha \beta} = \left(-1, +1, +1, +1\right)$ plus a metric perturbation $h_{\alpha\beta}$, 
\begin{equation}
	g_{\alpha \beta} = \eta_{\alpha \beta} + h_{\alpha\beta} \quad\quad \Longrightarrow \quad\quad 
	\overline{g}^{\alpha\beta} = \eta^{\alpha \beta} - \overline{h}^{\alpha\beta}\;,
\label{post_Newtonian_metric_5}
\end{equation}

\noindent
where $\overline{g}^{\alpha\beta} = \sqrt{-g}\,g^{\alpha\beta}$ are the contravariant components of the metric density, with $g = {\rm det}\left(g_{\mu\nu}\right)$ being 
the determinant of the metric. 
The decomposition (\ref{post_Newtonian_metric_5}) implies that the metric perturbations $h_{\alpha\beta}$ can be thought of as symmetric tensorial fields
which propagate in the flat background manifold ${\cal M}_0$. The metric of the flat background manifold is given by $\eta_{\alpha \beta}$. 
Thus, the flat background space-time is described by the pair $\left({\cal M}_0, \eta_{\mu\nu}\right)$, and the diffeomorphism between the physical manifold ${\cal M}$ and the 
flat background manifold ${\cal M}_0$ implies a one-to-one correspondence of the points ${\cal Q} \in {\cal M}_0$ to the points ${\cal P} \in {\cal M}$. 

The metric perturbation $h_{\alpha\beta}$ and the metric density perturbation $\overline{h}_{\alpha\beta}$ are uniquely related to each other:  
$h_{\alpha\beta} = \overline{h}_{\alpha\beta} - \frac{1}{2}\,\overline{h}\,\eta_{\alpha\beta}$ with $\overline{h} = \overline{h}^{\mu\nu}\eta_{\mu\nu}$.
The weak-field condition $\left|h_{\alpha\beta}\right| \ll 1$ inherits $|\overline{h}^{\alpha\beta}| \ll 1$.
In linearized gravity, the tensor indices are lowered and raised by the flat Minkowskian metric, e.g. $h^{\alpha\beta} = h_{\mu\nu}\,\eta^{\mu\alpha}\,\eta^{\mu\beta}$. 

Inserting (\ref{post_Newtonian_metric_5}) into the field equations of gravity and keeping terms linear in the metric perturbation, yields the field equations 
of linearized gravity (cf. Eq.~(18.5) in (Misner, Thorne, Wheeler, 1973)). They are considerably be simplified by the harmonic gauge, 
which implies that the coordinates $\{x\}$, which cover the flat background manifold ${\cal M}_0$, satisfy the equation $\square\,x^{\mu} = 0$. Then, 
the linearized field equations of gravity read 
\begin{equation}
	\square\,\overline{h}_{\alpha\beta} = - \frac{16 \pi G}{c^4}\,T_{\alpha\beta}\;,
\label{linearized_gravity}
\end{equation}

\noindent 
where $\square = \eta^{\mu\nu} \partial_{\mu} \partial_{\nu}$ is the flat d'Alembertian. Imposing Fock-Sommerfeld boundary conditions 
ensures a unique solution of (\ref{linearized_gravity}) in the coordinates $\{x\}$. Though, the harmonic gauge, $\square\,x^{\mu} = 0$, 
does not uniquely determine these coordinates, but allows for small deformations (Box $18.2$ in (Misner, Thorne, Wheeler, 1973) or Eq.~(3.521) 
in (Kopeikin, Efroimsky \& Kaplan, 2012)) 
\begin{equation}
        x^{\alpha}_{{\rm can}} = x^{\alpha} + \xi^{\alpha}(x^{\beta}) ,
\label{coordinate_transformation}
\end{equation}

\noindent
if the vector fields $\xi^{\alpha}$ satisfy $\square\,\xi^{\alpha} = 0$. The label of these new coordinates $\{x_{\rm can}\}$ abbreviates the term "canonical".  
The transformation (\ref{coordinate_transformation}) implies a transformation of the metric tensor,  
\begin{equation}
        g_{\alpha\beta}\left(t,\ve{x}\right)
        = \frac{\partial x^{\mu}_{\rm can}}{\partial x^{\alpha}}\, \frac{\partial x^{\nu}_{\rm can}}{\partial x^{\beta}}\,
	g_{\mu\nu}^{\rm can}\left(t_{\rm can} , \ve{x}_{\rm can}\right). 
\label{transformation_metric_tensor_A1}
\end{equation}

\noindent
By inserting (\ref{coordinate_transformation}) into (\ref{transformation_metric_tensor_A1}) and performing a series expansion
of the metric tensor on the r.h.s. around the old coordinates $\{x\}$, one obtains
(with notation $\partial_{\alpha}\,f \equiv f_{\,,\,\alpha} \equiv \partial f/ \partial x^{\alpha}$): 
\begin{equation}
	g_{\alpha\beta} \left(t,\ve{x}\right) = g_{\alpha\beta}^{\rm can} \left(t,\ve{x}\right) + \partial_{\alpha} \xi_{\beta} \left(t,\ve{x}\right)
        + \partial_{\beta} \xi_{\alpha}\left(t,\ve{x}\right), 
\label{transformation_metric_tensor_A2}
\end{equation}

\noindent
up to terms of higher order, i.e. up to non-linear terms. 
As stated above, by imposing the Fock-Sommerfeld boundary condition, the solution for the metric tensor $g_{\alpha\beta}$ in (\ref{transformation_metric_tensor_A2}) is unique. This 
unique solution can be expressed in terms of six Cartesian symmetric and tracefree (STF) multipoles $\{\hat{M}_L, \hat{S}_L, \hat{W}_L, \hat{X}_L, \hat{Y}_L, \hat{Z}_L\}$ (Thorne, 1980); 
the {\it hat} over the multipoles indicates STF. 
The canonical piece $g_{\alpha\beta}^{\rm can}$ in (\ref{transformation_metric_tensor_A2}) depends on two multipoles only: mass-multipoles and spin-multipoles $\{\hat{M}_L, \hat{S}_L\}$.
Accordingly, the gauge transformation of the metric tensor, as given by Eq.~(\ref{transformation_metric_tensor_A2}), results in the following form for the
metric perturbations ((Thorne, 1980) and (Blanchet \& Damour, 1986) and (Damour \& Iyer, 1991)): 
\begin{equation}
	h_{\alpha\beta}\left(t,\ve{x}\right) = h^{{\rm can}}_{\alpha\beta} \left[\hat{M}_L, \hat{S}_L\right]
	+ \partial_{\alpha} \xi_{\beta}\left[\hat{W}_L, \hat{X}_L, \hat{Y}_L, \hat{Z}_L\right] 
	+ \partial_{\beta} \xi_{\alpha}\left[\hat{W}_L, \hat{X}_L, \hat{Y}_L, \hat{Z}_L\right].
        \label{PN_Expansion_1PN_15PN}
\end{equation}

\noindent
In our investigation, the metric of the curved space-time in the exterior of the massive body is assumed to be time-independent.
Then, the canonical metric perturbations in (\ref{PN_Expansion_1PN_15PN}) are separated into two pieces,
$h^{\rm can}_{\alpha\beta} = h_{\alpha\beta}^{\left(2\right)\,{\rm can}} + h_{\alpha\beta}^{\left(3\right)\,{\rm can}}$, which are given by
\begin{equation}
        h_{00}^{\left(2\right)\,{\rm can}} = \frac{2}{c^2} \sum\limits_{l=0}^{\infty} \frac{\left(-1\right)^l}{l!}\,\hat{M}_L\,
	\hat{\partial}_L \frac{1}{r} \quad\quad {\rm and} \quad\quad  
        h_{0i}^{\left(3\right)\,{\rm can}} = \frac{4}{c^3} \sum\limits_{l=1}^{\infty} \frac{\left(-1\right)^l\,l}{\left(l+1\right)!} \,
        \epsilon_{iab}\,\hat{S}_{b L-1}\, \hat{\partial}_{a L-1} \frac{1}{r}\;,
        \label{Metric}
        \end{equation}

\noindent
while $h_{ij}^{\left(2\right)\,{\rm can}}  = h_{00}^{\left(2\right)\,{\rm can}}\,\delta_{ij}$ and  
the multipoles $\hat{M}_L$ and $\hat{S}_L$ are given by Eqs.~(5.33) and (5.35) in (Damour \& Iyer, 1991). 
The gauge functions in (\ref{PN_Expansion_1PN_15PN}) have been determined by (Thorne, 1980) and (Blanchet \& Damour, 1986) and (Damour \& Iyer, 1991) and read: 
\begin{equation}
	\xi^{0} = \sum\limits_{l=0}^{\infty} \hat{\partial}_L \frac{\hat{W}_L}{r} \quad \quad {\rm and} \quad \quad 
	\xi^{i} = \sum\limits_{l=0}^{\infty} \hat{\partial}_{i L}\,\frac{\hat{X}_L}{r} + \sum\limits_{l=1}^{\infty} \hat{\partial}_{L - 1}\,\frac{\hat{Y}_{i L - 1}}{r}
	+ \epsilon_{i ab}\,\sum\limits_{l=1}^{\infty} \hat{\partial}_{a L - 1}\,\frac{\hat{Z}_{b L - 1}}{r} \;.
        \label{gaugefunction}
\end{equation}

\noindent
Here, $r = \left|\ve{x}\right|$, and
\begin{equation}
        \hat{\partial}_L = {\rm STF}_{i_1 \dots i_l}\,\frac{\partial}{\partial x^{i_1}} \dots \frac{\partial}{\partial x^{i_l}}\,,
\label{Differential_Operator}
\end{equation}

\noindent
where the {\it hat} in $\hat{\partial}_L$ indicates STF operation with respect to the indices $L = i_1 \dots i_l$. The multipoles 
$\hat{W}_L, \hat{X}_L, \hat{Y}_L, \hat{Z}_L$ of the gauge functions in (\ref{gaugefunction}) are given in (Damour \& Iyer, 1991), 
but their explicit form is not relevant here, because we will show that the gauge terms in (\ref{PN_Expansion_1PN_15PN}) have no impact 
on the unit tangent vector and, therefore, no impact on the total light deflection. 
This result is an example of the general fact that $g_{\alpha\beta}$ and $g_{\alpha\beta}^{\rm can}$ in (\ref{transformation_metric_tensor_A2}) 
are physically equivalent, because they lead to same observables.


       \vspace*{0.7cm}
        \noindent {\large\bf 3. THE GEODESIC EQUATION}
        \smallskip
        
The light signal is assumed to propagate in the flat background manifold ${\cal M}_0$ which is covered by harmonic coordinates, $x^{\mu} = \left(x^0, x^1, x^2, x^3\right)$, 
where the origin of the spatial axes is located at the center of mass of the body. The exact light trajectory can be written in the following form, 
\begin{equation}
\ve{x}\left(t\right) = \ve{x}_0 + c \left(t-t_0\right) \ve{\sigma} + \Delta \ve{x}\left(t\right),  
\label{Introduction_2}
\end{equation}

\noindent
where $\Delta\ve{x}$ denotes the corrections to the unperturbed light trajectory, $\ve{x}_{\rm N}\left(t\right) = \ve{x}_0 + c \left(t-t_0\right) \ve{\sigma}$, and N stands for 
Newtonian (e.g. Kopeikin, Efroimsky \& Kaplan, 2012). Furthermore, we introduce the unit tangent vectors along the light trajectory at past and future infinity,
\begin{equation}
	\ve{\sigma} = \frac{\dot{\ve{x}}\left(t\right)}{c}\bigg|_{t \rightarrow - \infty} \quad\quad {\rm and} \quad\quad 
       \ve{\nu} = \frac{\dot{\ve{x}}\left(t\right)}{c}\,\bigg|_{t \rightarrow + \infty} \;,
        \label{vector_sigma_nu}
\end{equation}

\noindent
where a dot means total derivative with respect to coordinate time, and from (\ref{vector_sigma_nu}) follows $\ve{\sigma} \cdot \ve{\sigma} = 1$ and $\ve{\nu} \cdot \ve{\nu} = 1$. 
The total light deflection is the angle between these unit vectors,
\begin{equation}
        \delta\left(\ve{\sigma}, \ve{\nu}\right) = \arcsin \left| \ve{\sigma} \times \ve{\nu} \right| . 
        \label{Introduction_25}
\end{equation}

\noindent
The evaluation of this quantity is essential, in order to decide which multipoles need to be implemented in the relativistic model of light propagation for a given astrometric accuracy.

        \begin{figure}[h]
        \label{Diagram}
        \begin{center}
        \includegraphics[scale=0.175]{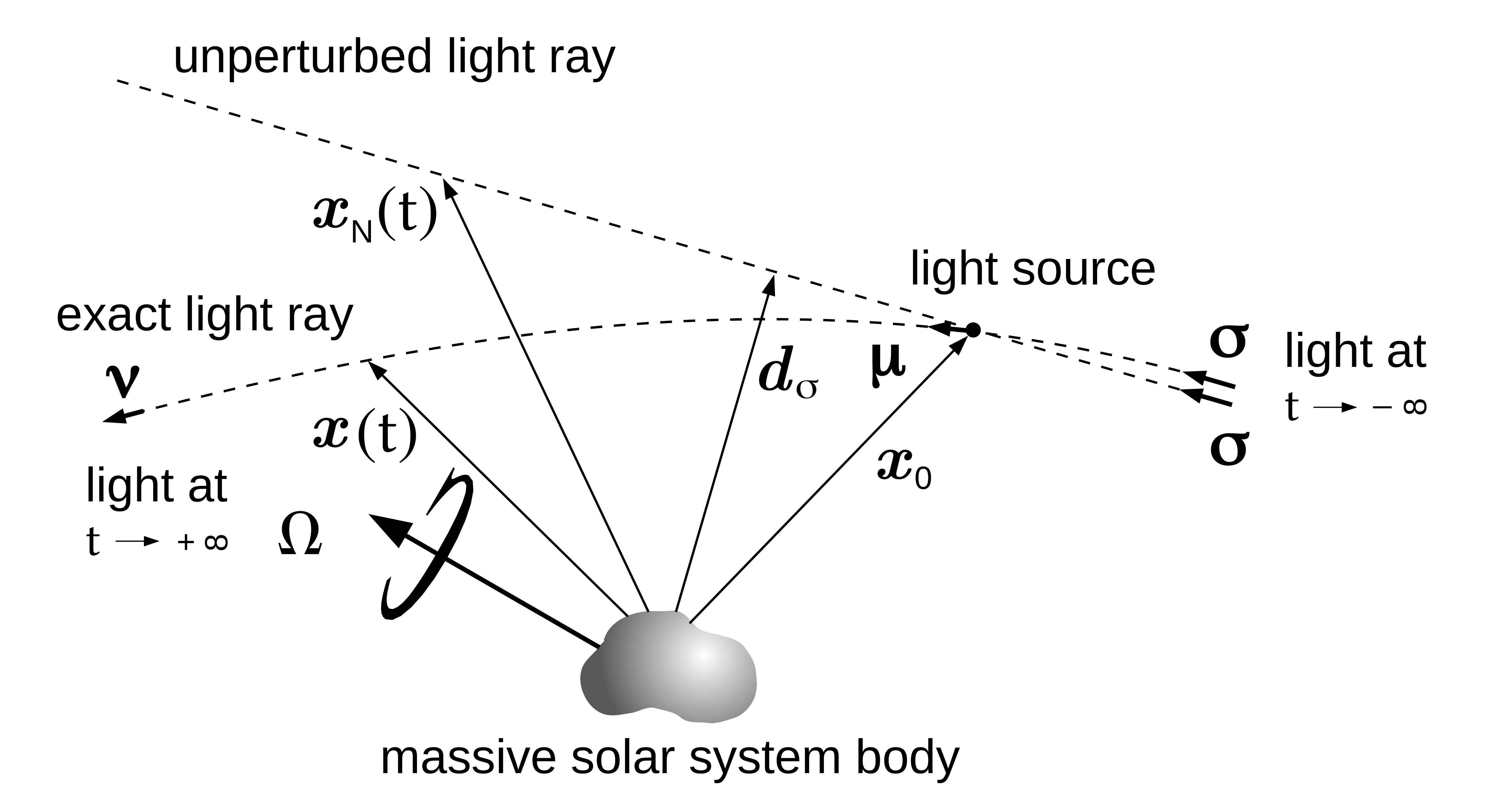}
        \caption{The light signal is emitted by the celestial light source at $\ve{x}_0$ in the direction of unit-vector $\ve{\mu}$ and propagates along the exact
        trajectory $\ve{x}\left(t\right)$. The origin of the spatial coordinates is located at the center of mass of the body, and the spatial coordinate axes 
	are aligned with the principal axes of the body. The body is in rotational motion around some axis with angular 
	velocity $\Omega$. The unit tangent vectors $\ve{\sigma}$ and $\ve{\nu}$ of the light trajectory at past infinity and future infinity are
	defined by Eqs.~(\ref{vector_sigma_nu}), while $\ve{d}_{\sigma}$ is the impact vector of the unperturbed light ray.}
        \end{center}
        \end{figure}

The geodesic equation for light rays in the post-Newtonian (PN) scheme in 1.5PN approximation reads (Kopeikin, Efroimsky \& Kaplan, 2012) 
(with notation $f_{\,,\,i} \equiv \partial f/ \partial x^i$):  
\begin{equation}
\frac{\ddot{x}^i \left(t\right)}{c^2} =
\frac{1}{2} \,h_{00,i}
- h_{00,j}\,\sigma^i \sigma^j - h_{ij,k}\, \sigma^j \sigma^k + \frac{1}{2}\,h_{jk,i}\,\sigma^j\sigma^k
- h_{0i,j}\,\sigma^j + h_{0j,i}\,\sigma^j - h_{0j,k}\,\sigma^i\,\sigma^j \sigma^k\;, 
\label{geodesic_equation_5}
\end{equation}

\noindent 
where the double-dot means twice the total derivative with respect to the coordinate time.  
Eq.~(\ref{geodesic_equation_5}) is valid up to terms of the post-post-Newtonian order ${\cal O}\left(c^{-4}\right)$, and
all those terms have been omitted which contain a derivative of the metric perturbations with respect to time, because we
consider the stationary case, that is the case of time-independent metric. Note, that in stationary case the geodesic equation in 1.5PN approximation 
in (\ref{geodesic_equation_5}) and the geodesic equation in 1PM approximation of the post-Minkowskian (PM) scheme agree with each other; 
cf. Eqs.~(A.4) and (A.6) in (Klioner \& Peip, 2003). 
If one inserts the metric perturbation (\ref{PN_Expansion_1PN_15PN}) into the geodesic equation (\ref{geodesic_equation_5}), 
one may separate the geodesic equations into a canonical term, $\ddot{\ve{x}}_{\rm can}$, plus a gauge term, $\ddot{\ve{x}}_{\rm gauge}$, as follows:
\begin{equation}
        \frac{\ddot{\ve{x}} \left(t\right)}{c^2} = \frac{\ddot{\ve{x}}_{\rm can} \left(t\right)}{c^2} + \frac{\ddot{\ve{x}}_{\rm gauge} \left(t\right)}{c^2}\;, 
\label{geodesic_equation_10}
\end{equation}

\noindent
where the spatial components of these terms are
\begin{eqnarray}
        \frac{\ddot{x}_{\rm can}^i \left(t\right)}{c^2} &=& h^{\left(2\right)\,{\rm can}}_{00,i} - 2\,h^{\left(2\right)\,{\rm can}}_{00,j}\,\sigma^i \sigma^j 
- h^{\left(3\right)\,{\rm can}}_{0i,j}\,\sigma^j + h^{\left(3\right)\,{\rm can}}_{0j,i}\,\sigma^j - h^{\left(3\right)\,{\rm can}}_{0j,k}\,\sigma^i\,\sigma^j \sigma^k\;, 
\label{geodesic_equation_can}
\\
        \frac{\ddot{x}_{\rm gauge}^i \left(t\right)}{c^2} &=& \partial_j\,\xi^0_{\,,\,k} \,\sigma^i \sigma^j \sigma^k - \partial_j\,\xi^i_{\,,\,k} \,\sigma^j \sigma^k \;. 
\label{geodesic_equation_gauge}
\end{eqnarray}

\noindent
The metric perturbations in (\ref{geodesic_equation_can}) are given by (\ref{Metric}), while the gauge functions in (\ref{geodesic_equation_gauge}) 
are given by (\ref{gaugefunction}); notice $\ve{x} = \ve{x}_{\rm N} + {\cal O}(c^{-2})$ and $r = |\ve{x}_{\rm N}| + {\cal O}(c^{-2})$ according to Eq.~(\ref{Introduction_2}).  
The first integration of (\ref{geodesic_equation_10}) yields the coordinate velocity of the light signal, 
\begin{equation}
	\frac{\dot{\ve{x}} \left(t\right)}{c} = \ve{\sigma} + \frac{\dot{\ve{x}}_{\rm can}\left(t\right)}{c} + \frac{\dot{\ve{x}}_{\rm gauge}\left(t\right)}{c}\;,  
\label{geodesic_equation_15}
\end{equation}

\noindent
and the unit tangent vectors (\ref{vector_sigma_nu}) are obtained from (\ref{geodesic_equation_15}) by taking the limit at plus and minus infinity. 
In the Appendix it is shown that the gauge terms (\ref{geodesic_equation_gauge}) do not contribute to these unit tangent vectors, 
because their first time derivative vanishes at plus and minus infinity, 
\begin{equation}
        \lim_{t \rightarrow \pm \infty}\frac{\dot{\ve{x}}_{\rm gauge} \left(t\right)}{c} = 0\;.
\label{first_integration_gauge}
\end{equation}

\noindent 
Accordingly, only the canonical terms in (\ref{geodesic_equation_can}) contribute to the unit tangent vector and, therefore, contribute to the total light deflection. 


       \vspace*{0.7cm}
       \noindent {\large\bf 4. TOTAL LIGHT DEFLECTION IN FIELD OF ARBITRARY BODY}
       \smallskip
        
As stated above, the gauge terms in (\ref{geodesic_equation_gauge}) do not contribute to the unit tangent vectors at plus and minus infinity (see Appendix),   
and there is no need to account for these terms. The first integration of the canonical terms (\ref{geodesic_equation_can}) in the geodesic equation has been performed 
in (Kopeikin, 1997). Taking the limit at plus infinity one arrives at the following expression for the unit tangent vector in (\ref{vector_sigma_nu}),   
\begin{equation}
        \ve{\nu} = \ve{\sigma} + \sum\limits_{l=0}^{\infty} \ve{\nu}_{\rm 1PN}^{M_L} + \sum\limits_{l=1}^{\infty} \ve{\nu}_{\rm 1.5PN}^{S_L} + {\cal O}\left(c^{-4}\right)\,. 
        \label{vector_mu_M_L_S_L}
\end{equation}

\noindent
The individual terms in (\ref{vector_mu_M_L_S_L}) are given by (limits of Eqs.~(34) and (37) in (Kopeikin, 1997)), 
\begin{eqnarray}
        {\nu}_{\rm 1PN}^{i\,M_L} &=& - \frac{4 G}{c^2}\,\frac{\left(-1\right)^l}{l!} \,\hat{M}_L 
        \,P^{ij} \frac{\partial}{\partial \xi^j}\,\hat{\partial}_L\,\ln |\ve{\xi}|, 
        \label{vector_mu_M_L_Simpler_Form}
        \\
        {\nu}_{\rm 1.5PN}^{i\,S_L} &=& - \frac{8 G}{c^3} \frac{\left(-1\right)^l}{l!} \frac{l}{l+1} \sigma^c \epsilon_{i_lbc} \, \hat{S}_{bL-1} 
        \, P^{ij} \frac{\partial}{\partial \xi^j}\,\hat{\partial}_{L}\,\ln |\ve{\xi}|\,,
        \label{vector_mu_S_L_Simpler_Form}
\end{eqnarray}

\noindent
where $P^{ij} = \delta^{ij} - \sigma^i \sigma^j$, and $\xi^i = P^i_j\,x_{\rm N}^j$ which will later be identified with the impact vector $\ve{d}_{\sigma}$
(cf. text below Eq.~(\ref{total_light_deflection_M_L_S_L})). The differential operator in (\ref{vector_mu_M_L_Simpler_Form}) and (\ref{vector_mu_S_L_Simpler_Form}) 
is given by (cf. Eq.~(24) in (Kopeikin, 1997) or Eq.~(30) in (Zschocke, 2022))
\begin{equation}
        \widehat{\partial}_{L} = {\rm STF}_{i_1 \dots i_l}\;\sum\limits_{p=0}^{l} \frac{l!}{\left(l-p\right)!\;p!}\;\sigma_{i_1}\,...\,\sigma_{i_p} \; 
        P_{i_{p+1}}^{j_{p+1}}\;...\;P_{i_l}^{j_l} 
        \;\frac{\partial}{\partial \xi^{j_{p+1}}}\;...\;
\frac{\partial}{\partial \xi^{j_{l}}}\;\left(\frac{\partial}{\partial c\tau}\right)^p \,.  
\label{Transformation_Derivative_3}
\end{equation}

\noindent
The operator (\ref{Differential_Operator}) is w.r.t. spatial coordinates $x^{a}$, while the operator (\ref{Transformation_Derivative_3}) is 
w.r.t. new variables $c \tau$ and $\xi^{a}$, and the notation {\it hat} in (\ref{Differential_Operator}) and {\it wide hat} in 
(\ref{Transformation_Derivative_3}) refers to this fact. 

Because $\ln \left|\ve{\xi}\right|$ in (\ref{vector_mu_M_L_Simpler_Form}) and (\ref{vector_mu_S_L_Simpler_Form}) is independent of variable $c \tau$, only the term 
$p=0$ in (\ref{Transformation_Derivative_3}) is relevant, which considerable simplifies the differential operator in (\ref{Transformation_Derivative_3}). 
A longer algebraic calculation leads finally to the following remarkable result (Zschocke, 2023):  
\begin{equation}
        \widehat{\partial}_{L}\,\ln \left|\ve{\xi}\right| = \frac{\left(-1\right)^{l+1}}{\left| \ve{\xi}\right|^l}\;{\rm STF}_{i_1 \dots i_l}\;  
        \sum\limits_{n=0}^{[l/2]} G_n^l 
        \,P_{i_1 i_2}\, \dots \, P_{i_{2 n - 1} i_{2 n}}\, \frac{\xi_{i_{2 n + 1}}\,\dots\,\xi_{i_{l}}}{\left|\ve{\xi}\right|^{l-2n}}\;,  
        \label{Relation_C}
\end{equation}

\noindent
which is valid for any natural number $l \ge 1$. The scalar coefficients in (\ref{Relation_C}) are given by
\begin{equation}
        G^l_n = \left(-1\right)^{n}\,2^{l - 2 n - 1}\,\frac{l!}{n!}\,\frac{\left(l - n - 1\right)!}{\left(l - 2 n\right)!}\;.  
        \label{Relation_D}
\end{equation}

\noindent
Remarkably, these coefficients coincide with the coefficients of the power series representation of Chebyshev polynomials of first kind $T_l$ in (\ref{Chebyshev_Polynomials}) 
up to a constant factor $(l-1)!\,$. In other words, the expression in (\ref{Relation_C}) is the generator of the coefficients of Chebyshev polynomials of first kind.


       \vspace*{0.7cm}
        \noindent {\large\bf 5. TOTAL LIGHT DEFLECTION IN FIELD OF AXISYMMETRIC BODY}
        \smallskip

In order to determine the mass-multipoles $\hat{M}_L$ and spin-multipoles $\hat{S}_L$ in (\ref{vector_sigma_nu}), the solar system bodies are described by a rigid axisymmetric 
structure and with arbitrary radial-dependent mass-density. Furthermore, the body is assumed to be in uniform rotational motion around its symmetry axis $\ve{e}_3$. For such an 
axisymmetric body the mass-multipoles and spin-multipoles have been calculated in (Zschocke, 2022) and depend on four physical parameters of the body: mass $M$, equatorial radius $P$, 
zonal harmonic coefficients $J_l$, angular velocity $\Omega$. Then, it has been shown in (Zschocke, 2023) that for such an axisymmetric body the mass-multipole and 
spin-multipole terms in (\ref{vector_mu_M_L_Simpler_Form}) are given by Chebyshev polynomials of first kind and second kind, 
\begin{eqnarray}
	\nu_{\rm 1PN}^{i\,M_L} &=& - \frac{4 G M}{c^2}\,\frac{J_l}{l}\,\left[1 - \left(\ve{\sigma} \cdot \ve{e}_3\right)^2\right]^{[l/2]} 
        P^{ij}\,\frac{\partial}{\partial \xi^j} \left(\frac{P}{\left|\ve{\xi}\right|}\right)^l\,T_l \left(x\right), 
        \label{Tangent_nu_M_Chebyshev}
	\\
	{\nu}_{\rm 1PN}^{i\,S_L} &=& - \frac{8 G M}{c^3}\,\Omega\,P\,\frac{J_{l-1}}{l+4}\,
        \left[1 - \left(\ve{\sigma} \cdot \ve{e}_3\right)^2\right]^{[l/2]}
        P^{ij}\,\frac{\partial}{\partial \xi^j} \frac{\left(\ve{\sigma} \times \ve{d}_{\sigma}\right) \cdot \ve{e}_3}{d_{\sigma}}
        \left(\frac{P}{\left|\ve{\xi}\right|}\right)^l\,U_{l-1} \left(x\right),
        \label{Tangent_nu_S_Chebyshev}
\end{eqnarray}

\noindent
where the power representations of the Chebyshev polynomials read (Arfken \& Weber, 1995),  
\begin{equation}
	T_l \left(x\right) = \frac{l}{2} \sum \limits_{n=0}^{[l/2]} \frac{\left(-1\right)^n}{n!} \,\frac{\left(l - n - 1\right)!}{\left(l - 2 n\right)!}
	\,\left(2 x\right)^{l - 2 n} \quad {\rm and} \quad 
	U_l \left(x\right) = \sum \limits_{n=0}^{[l/2]} \frac{\left(-1\right)^n}{n!} \,\frac{\left(l - n\right)!}{\left(l - 2 n\right)!}
        \,\left(2 x\right)^{l - 2 n}\;, 
        \label{Chebyshev_Polynomials}
\end{equation}

\noindent
with $T_0 = 1$. The real variable $x$ in (\ref{Tangent_nu_M_Chebyshev}) and (\ref{Tangent_nu_S_Chebyshev}) is defined by
\begin{equation}
x = \left(1 - \left(\ve{\sigma} \cdot \ve{e}_3\right)^2\right)^{-1/2}\;\left(\frac{\ve{d}_{\sigma} \cdot \ve{e}_3}{d_{\sigma}}\right) \quad \quad 
	{\rm where} \quad\quad - 1 \le x \le + 1\;. 
        \label{Variable_x_Chebyshev_A}
\end{equation}

\noindent 
It is just this highly remarkable fact, that the tangent vector $\ve{\nu}$ is given by Chebyshev polynomials, which allows
for a strict determination of the upper limits of the angle of total light deflection in (\ref{Introduction_25}). This is because the upper limits of Chebyshev polynomials are given by
\begin{equation}
        \left|T_l\right| \le 1 \quad {\rm and} \quad \left|U_{l-1}\right| \le l \;.
        \label{Introduction_Chebyshev_polynomials_2}
\end{equation}

\noindent
Accordingly, in the 1.5PN approximation the total light deflection (\ref{Introduction_25}) is given by
\begin{equation}
        \delta\left(\ve{\sigma},\ve{\nu}\right) = \sum\limits_{l=0}^{\infty} \delta\!\left(\ve{\sigma},\ve{\nu}_{\rm 1PN}^{M_L}\right) 
        + \sum\limits_{l=1}^{\infty} \delta\!\left(\ve{\sigma},\ve{\nu}_{\rm 1.5PN}^{S_L}\right). 
        \label{total_light_deflection}
\end{equation}

\noindent
The individual terms are given by ((Kopeikin, 1997), (Klioner, 1991), (Zschocke, 2023)) 
\begin{equation}
	\delta\left(\ve{\sigma},\ve{\nu}_{\rm 1PN}^{M_L}\right) = - \ve{\nu}_{\rm 1PN}^{M_L} \cdot \frac{\ve{d}_{\sigma}}{d_{\sigma}}   
	 \quad\quad {\rm and} \quad\quad 
         \delta\left(\ve{\sigma},\ve{\nu}_{\rm 1.5PN}^{S_L}\right) = - \ve{\nu}_{\rm 1.5PN}^{S_L} \cdot \frac{\ve{d}_{\sigma}}{d_{\sigma}}\;, 
         \label{total_light_deflection_M_L_S_L}
\end{equation}

\noindent
where $\ve{d}_{\sigma} = \ve{\sigma} \times \left(\ve{x}_0 \times \ve{\sigma}\right)$ is the impact vector,  
pointing from the body towards the unperturbed light ray at their closest distance. The absolute value, $d_{\sigma} = |\ve{d}_{\sigma}|$,
is the impact parameter. By inserting (\ref{Tangent_nu_M_Chebyshev}) and (\ref{Tangent_nu_S_Chebyshev}) into (\ref{total_light_deflection_M_L_S_L}) 
one obtains the following expressions for the individual mass-multipole and spin-multipole terms in the angle of total light deflection (\ref{total_light_deflection})
\begin{eqnarray}
	\delta\left(\ve{\sigma}, \ve{\nu}_{\rm 1PN}^{M_L}\right) &=& 
        - \frac{4 G M}{c^2 d_{\sigma}} J_l \left(\frac{P}{d_{\sigma}}\right)^l
	\left[1 - \left(\ve{\sigma} \cdot \ve{e}_3\right)^2\right]^{[l/2]} T_l\left(x\right), 
        \label{Total_Light_Deflection_Mass_Chebyshev}
	\\ 
	\delta\left(\ve{\sigma}, \ve{\nu}_{\rm 1.5PN}^{S_L}\right) &=&
        - \frac{8 G M}{c^3}\,J_{l-1} \frac{\Omega\;l}{l+4} \left(\frac{P}{d_{\sigma}}\right)^{l+1} 
        \frac{\left(\ve{\sigma} \times \ve{d}_{\sigma}\right) \cdot \ve{e}_3}{d_{\sigma}} 
        \left[1 - \left(\ve{\sigma} \cdot \ve{e}_3\right)^2\right]^{[l/2]}\,U_{l-1}\left(x\right)\!, 
        \label{Total_Light_Deflection_Spin_Chebyshev}
\end{eqnarray}

\noindent
where (\ref{Total_Light_Deflection_Mass_Chebyshev}) is valid for $l \ge 0$, while (\ref{Total_Light_Deflection_Spin_Chebyshev}) is valid for $l \ge 3$.
Thus far, it has not been possible to determine the upper limits of the total light deflection terms in (\ref{Total_Light_Deflection_Mass_Chebyshev})
and (\ref{Total_Light_Deflection_Spin_Chebyshev}), because these scalar functions are pretty much involved. In order to determine their upper limits, 
one actually would have to calculate their first derivatives with respect to variable $x$, and then to solve the corresponding algebraic equation  
of some order $n$, which is increasing with increasing multipole order $l$. However, according to the group theory of (Galois, 1846) there exist, in the general case,
no radicals for solving such equations for orders $n > 4$. Therefore, it is essential to recognize that the angle of 
total light deflection is just given in terms of Chebyshev polynomials of first and second kind. Only because of this important fact it is possible to determine the 
upper limits of (\ref{Total_Light_Deflection_Mass_Chebyshev}) and (\ref{Total_Light_Deflection_Spin_Chebyshev}) by means of relations (\ref{Introduction_Chebyshev_polynomials_2}).
Because the impact parameter is larger or equal to the equatorial radius of the body, $d_{\sigma} \ge P$,
one obtains from (\ref{Total_Light_Deflection_Mass_Chebyshev}) and (\ref{Total_Light_Deflection_Spin_Chebyshev}),
\begin{equation}
	\left|\delta\left(\ve{\sigma}, \ve{\nu}_{\rm 1PN}^{M_L}\right)\right| \le 
	\frac{4 G M}{c^2}\,\frac{\left|J_l\right|}{P} \quad\quad {\rm and} \quad\quad 
        \left|\delta\left(\ve{\sigma}, \ve{\nu}_{\rm 1.5PN}^{S_L}\right) \right| \le
        \frac{8 G M}{c^3}\,\Omega\,\frac{l^2}{l+4}\,\left|J_{l-1}\right|,
        \label{Summary_M_L_S_L_Grazing}
\end{equation}

\noindent
where the inequality on the l.h.s. and r.h.s. are valid for $l \ge 0$ and $l \ge 3$, respectively; for the case of spin-dipole ($l = 1$) one finds 
$\displaystyle \left|\delta\left(\ve{\sigma}, \ve{\nu}_{\rm 1.5PN}^{S_1}\right)\right| \le \frac{4 G M}{c^3}\,\Omega\,\kappa^2$ (Klioner, 1991). 
These inequalities (\ref{Summary_M_L_S_L_Grazing}) for the total light deflection are strictly valid in the 1PN and 1.5PN, and can be used to decide, 
whether a specific multipole term needs to be taken into account in the light propagation model for a given goal accuracy of future astrometry missions 
aiming at the sub-micro-arcsecond and nano-arcsecond level.
Some numerical values are presented in Table~\ref{Table1} for the case of light deflection of the giant planets Jupiter and Saturn.
\begin{table}[h!]
\caption{The upper limits of total light deflection at giant planets Jupiter and Saturn caused by their mass-multipoles and spin-multipoles according to
        Eqs.~(\ref{Summary_M_L_S_L_Grazing}).
        All values are given in micro-arcsecond ($\muas$). A blank entry indicates the light deflection is smaller than a nano-arcsecond (nas).
        For the physical parameters $M, P, J_l, \Omega$ standard values are used (Zschocke, 2023).}
\begin{tabular}{| c | c | c| | c | c | c|}
\hline
	&&&&&\\[-12pt]
Light deflection &\hbox to 20mm{\hfill Jupiter\hfill} &\hbox to 20mm{\hfill Saturn\hfill} & Light deflection &\hbox to 20mm{\hfill Jupiter\hfill} &\hbox to 20mm{\hfill Saturn\hfill}\\[3pt]
\hline
	&&&&&\\[-12pt]
$|\delta(\ve{\sigma}, \ve{\nu}_{\rm 1PN}^{M_0})|$ & $16.3 \times 10^{3}$ & $5.8 \times 10^{3}$ & $|\delta(\ve{\sigma}, \ve{\nu}_{\rm 1.5PN}^{S_1})|$ & $0.17$ & $0.04$  \\[3pt]
	$|\delta(\ve{\sigma}, \ve{\nu}_{\rm 1PN}^{M_2})|$ & $ 239 $ & $ 94$ & $|\delta(\ve{\sigma}, \ve{\nu}_{\rm 1.5PN}^{S_3})|$ & $ 0.026 $ & $ 0.008$ \\[3pt]
	$|\delta(\ve{\sigma}, \ve{\nu}_{\rm 1PN}^{M_4})|$ & $ 9.6 $ & $ 5.41 $ & $|\delta(\ve{\sigma}, \ve{\nu}_{\rm 1.5PN}^{S_5})|$ & $ 0.001 $ & $ - $ \\[3pt]
	$|\delta(\ve{\sigma}, \ve{\nu}_{\rm 1PN}^{M_6})|$ & $ 0.55 $ & $ 0.50 $ & $|\delta(\ve{\sigma}, \ve{\nu}_{\rm 1.5PN}^{S_7})|$ & $ - $ & $ - $ \\[3pt]
	$|\delta(\ve{\sigma}, \ve{\nu}_{\rm 1PN}^{M_8})|$ & $0.04 $ & $ 0.06 $ &  $|\delta(\ve{\sigma}, \ve{\nu}_{\rm 1.5PN}^{S_9})|$ & $ - $ & $ - $ \\[3pt]
	$|\delta(\ve{\sigma}, \ve{\nu}_{\rm 1PN}^{M_{10}})|$ & $0.003 $ & $ 0.01 $ & $|\delta(\ve{\sigma}, \ve{\nu}_{\rm 1.5PN}^{S_{11}})|$ & $ - $ & $ - $ \\[3pt]
\hline
\end{tabular}
\label{Table1}
\end{table}

	\vspace*{0.7cm}
	\noindent {\large\bf 6. CONCLUSION}
	\smallskip
	

The determination of the upper limits of the angle of total light deflection provides a criterion, up to which order in $l$ the mass-multipoles $\hat{M}_L$ and 
the spin-multipoles $\hat{S}_L$ need to be taken 
into account. Such a criterion simplifies considerably the relativistic modeling of light trajectories for future ultra-high precision astrometry missions on the sub-$\muas$ level 
of accuracy. In our investigation we have determined the unit tangent vector of the light ray at future infinity of the light trajectory by Eqs.~(\ref{Tangent_nu_M_Chebyshev}) and 
(\ref{Tangent_nu_S_Chebyshev}) as well as strict upper limits for the total light deflection angle by Eqs.~(\ref{Summary_M_L_S_L_Grazing}) for higher 
mass-multipoles and spin-multipoles. The remarkable fact, that the unit tangent vector of the light ray at future infinity is naturally given by Chebyshev polynomials, 
allows for a strict mathematical statement about the upper limits of the total light deflection.

        \vspace*{0.7cm}
        \noindent {\large\bf ACKNOWLEDGMENT}
        \smallskip

This work was funded by Deutsche Forschungsgemeinschaft (DFG): grant number 447922800.


\newpage 

       \vspace*{0.0cm}
        \noindent {\large\bf APPENDIX}
        \smallskip

In this appendix we will demonstrate the limit (\ref{first_integration_gauge}). 
The gauge terms in the geodesic equation (\ref{geodesic_equation_gauge}) consist of two pieces, 
$\ddot{\ve{x}}_{\rm gauge} = \ddot{\ve{x}}_{\rm g1} + \ddot{\ve{x}}_{\rm g2}$. Their spatial components are given by
\begin{equation}
	\frac{\ddot{x}_{\rm g1}^i \left(t\right)}{c^2} = +\, \partial_j\,\xi^0_{\,,\,k} \,\sigma^i \sigma^j \sigma^k \quad\quad {\rm and} \quad\quad   
        \frac{\ddot{x}_{\rm g2}^i \left(t\right)}{c^2} = -\, \partial_j\,\xi^i_{\,,\,k} \,\sigma^j \sigma^k\;,
        \label{Term}
\end{equation}

\noindent
where the gauge vectors are given by Eqs.~(\ref{gaugefunction}). Let us consider the first term in (\ref{Term}). Using 
$\left(r^{-1}\right)_{\,,\,jk} = 3 x_j x_k / r^5 - \delta_{jk} / r^3$, one obtains 
\begin{equation}
	\frac{\ddot{\ve{x}}_{\rm g1} \left(t\right)}{c^2} = + 2 \sum\limits_{l=0}^{\infty} \hat{\partial}_L\,\frac{\hat{W}_L}{r^3}\,\ve{\sigma} 
	- 3 \sum\limits_{l=0}^{\infty} \hat{\partial}_L\,\frac{\hat{W}_L}{r^5}\,\left(d_{\sigma}\right)^2\,\ve{\sigma} \;, 
\label{appendix_15}
\end{equation}

\noindent
where $\left(\ve{\sigma} \cdot \ve{x}\right)^2 = r^2 - \left(d_{\sigma}\right)^2$ has been used.
This expression has to be integrated over the time variable. To apply the advanced integration methods developed by (Kopeikin, 1997), we have to transform
(\ref{appendix_15}) from $\left(c t, \ve{x}\right)$ into terms of two new variables, $c \tau = \ve{\sigma} \cdot \ve{x}_{\rm N}$ and $\xi^i = P^{ij}\,x_{\rm N}^j$, which are independent of each other,  
and obtain (note that $\ve{\xi} = \ve{d}_{\sigma}$ hence $\left(d_{\sigma}\right)^2 = \ve{\xi} \cdot \ve{\xi} = \xi^2$)
\begin{equation}
	\frac{\ddot{\ve{x}}_{\rm g1} \left(\tau\right)}{c^2} = \sum\limits_{l=0}^{\infty} \hat{W}_L\,\widehat{\partial}_L  
        \left(\frac{2}{\left(\sqrt{\xi^2 + c^2 \tau^2}\right)^3} - \frac{3\;\left(\xi\right)^2}{\left(\sqrt{\xi^2 + c^2 \tau^2}\right)^5} \right) \, \ve{\sigma}\;, 
\label{appendix_20}
\end{equation}

\noindent
where the double-dot in (\ref{appendix_20}) means twice the total derivative with respect to variable $\tau$.
The  differential operator (\ref{appendix_20}) has been given by Eq.~(\ref{Transformation_Derivative_3}). 
To get the coordinate velocity of the light signal, one has to integrate (\ref{appendix_20}) over variable $c \tau$ and obtains for the spatial components  
\begin{equation}
	\frac{\dot{x}^i_{\rm g1} \left(\tau\right)}{c} = - \sum\limits_{l=0}^{\infty} \hat{W}_L\,\widehat{\partial}_L \,\frac{c \tau}{r^3}\;\sigma^i 
	= \frac{\partial}{\partial c \tau}\sum\limits_{l=0}^{\infty} \hat{W}_L\,\widehat{\partial}_L \,\frac{1}{r}\;\sigma^i\;.  
\label{appendix_30}
\end{equation}

\noindent
A similar calculation can be performed for the second gauge term in (\ref{Term}), which yields   
\begin{equation}
	\frac{\dot{x}^{i}_{\rm g2} \left(\tau\right)}{c} = \frac{\partial}{\partial c \tau} \sum\limits_{l=0}^{\infty} \hat{X}_{L}\,\widehat{\partial}_{iL}\,\frac{1}{r}
	+ \frac{\partial}{\partial c \tau} \sum\limits_{l=1}^{\infty} \hat{Y}_{iL}\,\widehat{\partial}_{L-1}\,\frac{1}{r}
	+ \epsilon_{iab} \frac{\partial}{\partial c \tau} \sum\limits_{l=1}^{\infty} \hat{Z}_{bL-1}\,\widehat{\partial}_{a L-1}\,\frac{1}{r}\;.
\label{appendix_35}
\end{equation}

\noindent 
By inserting (\ref{Transformation_Derivative_3}) into (\ref{appendix_30}) and (\ref{appendix_35}) one finds that these terms vanish at infinity, and we get  
\begin{equation}
\lim_ {\tau = \pm \infty} \frac{\dot{\ve{x}}_{\rm gauge} \left(\tau\right)}{c} = 
	\lim_ {\tau = \pm \infty} \frac{\dot{\ve{x}}_{\rm g1} \left(\tau\right)}{c} + \lim_ {\tau = \pm \infty} \frac{\dot{\ve{x}}_{\rm g2} \left(\tau\right)}{c} = 0\;.
\label{appendix_40}
\end{equation}

\noindent
Thus, by transforming (\ref{appendix_40}) back from $\left(c \tau, \ve{\xi}\right)$ into $\left(c t, \ve{x}\right)$,
we have shown the validity of (\ref{first_integration_gauge}).

\newpage 

	\vspace*{0.0cm}
	\noindent{\large\bf 7. REFERENCES}
	{
	
	\leftskip=5mm
	\parindent=-5mm
	\smallskip

Arfken, G.B., Weber, H.J., 1995,
``Mathematical methods for physicists'',
London, Academic Press, 4th Edition, 1995.

Blanchet, L., Damour, T., 1986,
``Radiative gravitational fields in general relativity: I. General structure of the field outside the source'',
Phil. Trans. R. Soc. London A, 320, pp. 379 - 430.

Damour, T., Iyer, B.R., 1991,
``Multipole analysis for electromagnetism and linearized gravity with irreducible Cartesian tensors'',
Phys. Rev. D, 43, pp. 3259 - 3272.

Einstein, A., 1915,
``Die Feldgleichungen der Gravitation'',
Sitz.ber. Akad. Wiss. Berlin, 2, pp. 844 - 847.

Galois, E., 1846, 
``M\'{e}moire sur les conditions de r\'{e}solubilit\'{e} des \'{e}quations par radicaux'', 
Journal de Math\'{e}matiques Pures et Appliqu\'{e}es, 11, pp. 381 - 444.  

Johnston, K.J., McCarthy, D.D., Luzum, B.J., Kaplan, G.H., 2000,
``Towards Models and Constants for Sub-Microarcsecond Astrometry'',
Proceedings of IAU Colloquium 180, U.S. Naval Observatory, Washington D.C., March 26 - April 2, 2000.

Klioner, S.A., 1991,
``Influence of the quadrupole field and rotation of objects on light propagation'',
Sov. Astron., 35, pp. 523 - 530.

Klioner, S.A., Peip, M., 
``Numerical simulations of the light propagation in the gravitational field of moving bodies'', 
A \& A, 410, pp. 1063 - 1074. 

Kopeikin, S.M., 1997,
``Propagation of light in the stationary field of multipole gravitational lens'',
J. Math. Phys., 38, pp. 2587 - 2601.

Kopeikin, S., Efroimsky, M., Kaplan, G., 2012,
``Relativistic Celestial Mechanics of the Solar System'',
Wiley-VCH, Signapure.

Misner, C.W., Thorne, K.S., Wheeler, J.A., 1973,
``Gravitation'', New York: W.H. Freeman. 

Thorne, K.S., 1980, 
``Multipole expansions of gravitational radiation'',
Rev. Mod. Phys., 52, pp. 299 - 339.

Zschocke, S., 2022,
``Time delay in the quadrupole field of a body at rest in 2PN approximation'',
Phys. Rev. D, 106, 104052, pp. 1 - 35.

Zschocke, S., 2023,
``Total light deflection in the gravitational field of an axisymmetric body at rest with full mass and spin multipole structure'',
Phys. Rev. D, 107, 124055, pp. 1 - 25.
	
\end{document}